\documentclass[11pt]{article}

\usepackage[margin=1in]{geometry}
\usepackage{setspace}
\usepackage{amsmath,amssymb,amsthm}
\usepackage{authblk}
\usepackage{booktabs}
\usepackage{graphicx}
\usepackage{float}
\usepackage{subcaption}
\usepackage{url}

\usepackage[authoryear]{natbib}

\newtheorem{theorem}{Theorem}[section]

\newtheorem{corollary}[theorem]{Corollary}
\newtheorem{conjecture}[theorem]{Conjecture}
\newtheorem{remark}[theorem]{Remark}

\title{\bf Leave-one-out testing for node-level differences\\
in Gaussian graphical models}

\author[1]{Davide Benussi\thanks{davide.benussi@phd.unipd.it}}
\author[2]{Ester Alongi\thanks{ester.alongi@phd.unipd.it}}
\author[1]{Erika Banzato\thanks{erika.banzato@unipd.it}}

\affil[1]{Department of Statistical Sciences, University of Padova, 35121, Padova, Italy}

\date{\today}

\begin{document}
\maketitle

\begin{abstract}
We study two-sample equality testing in Gaussian graphical models with the aim of localizing distributional differences at the node level. The proposed procedure targets departures in the joint distribution, which may arise from mean shifts, variance changes, or covariance differences.
We extend Bartlett-adjusted likelihood ratio tests to a leave-one-out framework on a fully connected working graph.
The resulting increments have standard chi-square null limits, enabling calibrated inference for single nodes and fixed-size subsets.
We also discuss how decomposable graphical structure can be exploited to apply the approach locally on larger graphs.
Simulation results indicate good calibration and competitive power, and a real data application highlights the practical applicability of the method.

\end{abstract}

\section{Introduction}\label{sec:Introduction}

Gaussian graphical models (GGMs) parametrize multivariate normal distributions by a mean vector and a precision matrix constrained by an undirected graph, where zeros in the precision encode conditional independences \citep{lauritzen1996graphical,whittaker2009graphical}. Methodological developments on GGMs have progressed mainly along three directions: structure learning, edge-wise inference, and equality testing between two conditions. This work focuses on the latter.

In decomposable graphs, the likelihood ratio statistic (LRT; \citealp{WILKS_1938}) admits convenient clique-separator factorizations and asymptotic chi-square limits, with finite-sample accuracy improved by Bartlett corrections \citep{bartlett1937properties,ERIKA_BANZATO_2023}. 
Recent contributions exploit decomposable structure to localize changes through clique-wise testing and set operations \citep{salviato2019sourceset,DJORDJILOVIC_CHIOGNA_2022}, but do not directly address node-level inference and rely on decomposability. By contrast, assuming a fully connected graph avoids this restriction and is often more realistic when the graph structure is unknown. 

We propose a node-level inference procedure for localizing general distributional differences between two GGMs. The aim is not to compare conditional independence structures themselves, but to identify nodes that contribute to a departure from equality of distributions. Such departures may arise from mean shifts, variance changes, or more general covariance differences, even when the underlying graph remains unchanged. Although it does not directly target the identification of zero off-diagonal entries in the precision matrix, the proposed approach can also localize changes in the dependence structure through the identification of nodes adjacent to perturbed edges.
Our procedure extends the two-sample Bartlett-adjusted LRT of \citet{ERIKA_BANZATO_2023} to a leave-one-out framework on a fully connected graph, either treated as a maximal clique or as a simplifying working assumption when the structure is unknown. While the fully connected graph corresponds to the saturated Gaussian graphical model, the proposed procedure is still rooted in the graphical modeling perspective. In particular, it can be seen as a node-level refinement of likelihood-based inference, extending classical clique-wise approaches to a finer resolution \citep{salviato2019sourceset}. For any candidate node or fixed-size subset, we compare the Bartlett-adjusted LRT on the full graph with the statistic recomputed after removing that node or subset. The resulting increment has, under the null hypothesis of equality of distributions, an asymptotic chi-square distribution with degrees of freedom depending only on the number of removed nodes and the total number of nodes. This yields calibrated node- and subset-level \(p\)-values while retaining the finite-sample accuracy of the Bartlett correction.

The remainder of the paper is organized as follows. Section~\ref{sec:proposal} introduces the leave-one-out statistics and establishes their null limiting distributions. Section~\ref{sec:Simulation study} presents simulation results for a fully connected graph, evaluating the validity of the proposed procedure. Section~\ref{sec:Extension to decomposable graphical models and scalability to larger graphs} describes how the proposed approach can be extended to decomposable graphical models and demonstrates how this framework facilitates scalability to larger graphs through localized low-dimensional inference. Section~\ref{sec:Conclusions} summarizes the main contributions and discusses limitations and possible directions for future research. Appendix~\ref{app:Appendix_A} provides proofs of the main results, along with additional simulation studies, including cases with localized edge-level perturbations, scenarios in which a sparse graph is treated as fully connected, and an application to real data.

\section{Our proposal}\label{sec:proposal}

Let \(G=(V,E)\) be a fully connected undirected graph with node set \(V=\{1,\ldots,p\}\), edge set \(E=\{(u,v):u,v\in V,\ u\neq v\}\), and let \(p=|V|\), where \(|\cdot|\) denotes set cardinality.
Suppose \(X_V\sim\mathcal N_p(\mu,\Sigma)\) is a \(p\)-variate normal vector whose components are indexed by the nodes in \(V\).
For \(c=1,2\), let \(X_{V,\iota}^{(c)}\sim\mathcal N_p(\mu^{(c)},\Sigma^{(c)})\) for \(\iota=1,\ldots,n_c\), and set \(\mathbb{X}_{V}^{(c)}=[X_{V,1}^{(c)},\ldots,X_{V,n_c}^{(c)}]^\top\in\mathbb R^{n_c\times p}\). The two samples are independent, with \(n=n_1+n_2\).
We test the global null
\begin{align}
H_0:\ (\mu^{(1)},\Sigma^{(1)})=(\mu^{(2)},\Sigma^{(2)}) 
\quad\text{vs}\quad 
H_1:\ \text{$H_0$ not true}, 
\label{global_H0}
\end{align}
restricting attention, under \(H_1\), to mean shifts and/or variance changes, i.e., \(\mu^{(1)}\neq\mu^{(2)}\) and/or \(\operatorname{diag}(\Sigma^{(1)})\neq \operatorname{diag}(\Sigma^{(2)})\).
Rejection of \(H_0\) indicates the presence of a distributional difference between the two groups, which may arise from differences in the mean vectors, the covariance matrices, or both. Our objective is to identify the nodes that contribute to the overall departure from \(H_0\).
The global LRT for \eqref{global_H0} is
\[
W_n^{V}
=\sum_{c=1}^2 n_c\log\frac{\det\widehat\Sigma}{\det\widehat\Sigma^{(c)}},
\]
where \(\widehat\Sigma\) is the MLE of \(\Sigma\) under \(H_0\) and \(\widehat\Sigma^{(c)}\) is the MLE in group \(c\). As \(n=n_1+n_2\to\infty\), \(W_n^{V}\overset{d}{\to}\chi^2_{f_V}\) with \(f_V=|V|(|V|+3)/2=p(p+3)/2\). A Bartlett-type correction improves finite-sample accuracy \citep{bartlett1937properties,ERIKA_BANZATO_2023}. Let
\[
r_{V,x}=\sqrt{-\log(1-|V|/x)},\qquad n_c'=n_c-1,
\]
and define
{\small{
\begin{align*}
\mu_n^{V}
=\frac{1}{4}\!\left[
-4|V|-\sum_{c=1}^2\frac{|V|}{n_c}
+ n\,{r^2_{V,n}}(2|V|-2n+3)
- \sum_{c=1}^2 n_c\, {r^2_{V,{n'_c}}}(2|V|-2n_c+3)\right]\!.
\end{align*}
}}
The adjusted test ststistic is
\[
T_n^{V}=\delta_n^{V}W_n^{V},\qquad 
\delta_n^{V}=\frac{f_V}{-2\mu_n^{V}},
\]
so that \(T_n^{V}\) is accurately approximated under \(H_0\) by a chi-square distribution with \(f_V\) degrees of freedom even when \(n\) and \(p\) are unbalanced \citep{ERIKA_BANZATO_2023}. The quantity \(\mu_n^{V}\) provides an approximation of the expected value of the likelihood ratio statistic \(W_n^{V}\) under the null hypothesis, as a function of the sample sizes and the dimension \(p\). 
The scaling factor \(\delta_n^{V}\) is therefore constructed to match the mean of the adjusted statistic \(T_n^{V}\) with the degrees of freedom \(f_V\) of its asymptotic chi-square distribution. 
In this way, \(T_n^{V}\) can be interpreted as a Bartlett-type corrected version of \(W_n^{V}\), designed to improve the accuracy of the chi-square approximation in finite samples, particularly when \(p\) is not negligible relative to \(n\).

To localize departures from \(H_0\) at the node or subset level, we compare the global evidence against equality of distributions before and after removing a given node or subset of nodes. Under the alternative hypothesis, let \(\mathcal{S}\subset V\) denote the unknown set of nodes responsible for the difference in distribution between the two groups, and write \(\mathcal{S}^c=V\setminus\mathcal{S}\) for the set of unaltered nodes. For any nonempty \(M\subset V\) with \(1 \le|M|\le p-1\), let \(V\setminus M\) be the node set of the induced subgraph and define \(T_n^{V\setminus M}\) analogously. The Bartlett-adjusted increment
\[
\Delta_{M}=T_n^{V}-T_n^{V\setminus M},\qquad 1\le |M|\le p-1,
\]
measures the increment in global LRT after removing \(M\). 
The proposed increments aim to localize contributions to general distributional differences.
\(\Delta_{M}\) is stochastically large when \(M\cap\mathcal S\neq\varnothing\) and asymptotically negligible when \(M\subset\mathcal S^c\). For a fixed cardinality \(l = |M|\), \(1\le l\le p-1\), let \(\mathcal{M}(l)=\{M\subset V:\ |M|=l\}\) be the set of all subsets of \(V\) with fixed cardinality \(l\).
The next result gives the null limit of \(\Delta_M\) and extends Theorem 1 of \cite{DJORDJILOVIC_CHIOGNA_2022}, originally stated for clique-level tests in decomposable graphs.

{
\renewcommand{\thetheorem}{1}
\begin{theorem}[Null limiting distribution of $\Delta_M$]\label{thm:null_distrib_delta_M}
Fix \(1\le l\le p-1\) and \(M\in\mathcal{M}(l)\). Assume the regularity conditions in
\citet{ERIKA_BANZATO_2023} ensuring chi-square limits for \(T_n^{V}\) and \(T_n^{V\setminus M}\)
hold. Under \(H_0\), as \(n=n_1+n_2\to\infty\),
\[
\Delta_M = T_n^{V}-T_n^{V\setminus M}\ \overset{d}{\longrightarrow}\ \chi^2_{\,h(l,p)},
\]
where
\[
h(l,p)=\frac{l(2p-l+3)}{2}.
\]
\end{theorem}
}

\begin{proof}
See  Appendix \ref{app:Appendix_A}.
\end{proof}

The regularity conditions on $n,n_1,n_2$ and $|V|,|V\setminus M|$ required by Theorem~\ref{thm:null_distrib_delta_M} coincide with those of \citet[Theorem~1]{ERIKA_BANZATO_2023}: 
$(p_n)_{n\in\mathbb{N}}$ is a sequence of integers with $1 \leq p_n < n_c-1$, $\min_{c=1,2} n_c \to \infty$, $|V|/n \to 0$, and $|V\setminus M|/n \to 0$. 
These conditions have a natural interpretation. The constraint $p_n < n_c - 1$ ensures that the sample covariance matrices are invertible, so that the likelihood ratio statistic is well defined. The requirement $\min_{c=1,2} n_c \to \infty$ guarantees that each group contains enough observations for asymptotic arguments to apply. Finally, the conditions $|V|/n \to 0$ and $|V\setminus M|/n \to 0$ ensure that the dimension of the problem grows slower than the sample size, which is necessary for the chi-square approximation of the (adjusted) likelihood ratio statistics to remain accurate.
The proof presented in Appendix~\ref{app:Appendix_A} then follows directly; see also Appendix~A of the Supplementary Material in \citealp{ERIKA_BANZATO_2023}. Note that $h(1,p)=p+1$, $h(2,p)=2p+1$, and $h(3,p)=3p$, corresponding to singletons, pairs, and triplets, which are our main focus, though Theorem~\ref{thm:null_distrib_delta_M} holds for any $1\le l \le p-1$.

{
\renewcommand{\thetheorem}{1}
\begin{corollary}[Singleton removal]\label{corollary:null_distrib_delta_j}
Let \(l=1\), \(j\in V\) and \(p\ge2\). Assume the regularity conditions of
Theorem~\ref{thm:null_distrib_delta_M}. Under \(H_0\), as \(n=n_1+n_2\to\infty\),
\[
\Delta_j \;=\; T_n^{V}-T_n^{V\setminus\{j\}} \overset{d}{\longrightarrow} \chi^2_{\,p+1}.
\]
\end{corollary}
}

Corollary~\ref{corollary:null_distrib_delta_j} is a direct consequence of Theorem~\ref{thm:null_distrib_delta_M} with \(l=1\).

{
\renewcommand{\thetheorem}{1}
\begin{remark}
(i) For fixed \(l\), the family \(\{\Delta_M: M\in\mathcal{M}(l)\}\) is identically distributed under \(H_0\), but generally dependent across overlapping \(M\).
(ii) Using \(W_n\) in place of \(T_n\) preserves the form of the results but degrades the finite-sample chi-square approximation.
(iii) Numerical experiments in Section \ref{sec:Simulation study} indicate \(W_n\)-based tests are anti-conservative, whereas \(T_n\)-based tests track the nominal level more closely.
\end{remark}
}

To assess whether \(M \in \mathcal{S}\), it suffices to know the asymptotic null distribution of \(\Delta_M\). Indeed, by standard large-sample theory, we test \(H_0:\Delta_M=0\) for a given \(M \in \mathcal{M}(l)\) by the marginal \(p\)-value
\begin{align*}
p_M = \Pr\big( \Delta_M \geq \Delta_M^{\mathrm{obs}} \,\big|\, H_0 \big), \label{p-value-computation}
\end{align*}
where \(\Delta_M\) under \(H_0\) is approximated by its chi-square limit and \(\Delta_M^{\mathrm{obs}}\) is the observed statistic.
Because \(|\mathcal{M}(l)|\) grows combinatorially, testing all \(\{\Delta_M:M\in\mathcal{M}(l)\}\) without adjustment inflates the probability of spurious findings, especially since the tests are dependent. We therefore apply Holm's step-down Bonferroni procedure \citep{bonferroni1936, Holm1979}, which controls the family-wise error rate (FWER) under arbitrary dependence when applied to valid null $p$-values. Procedures targeting the false discovery rate may offer greater power. For example, the Benjamini--Hochberg procedure \citep{Benjamini_Hochberg_1995} requires independence or suitable dependence conditions, which have not been established here, whereas the Benjamini--Yekutieli procedure \citep{Benjamini_Yekutieli_2001} accommodates arbitrary dependence but can also be conservative. We focus on FWER control, using the asymptotic calibration established under $H_0$ in Theorem~\ref{thm:null_distrib_delta_M}. For \(l=1\), this yields a node-level selection set, \(\widehat{\mathcal{S}}\), consisting of all nodes whose adjusted \(p\)-value is at most \(\alpha\), the nominal level. 
Theorem \ref{thm:null_distrib_delta_M} holds for any \(l\), but it suffices to consider \(l=1\) to obtain \(\widehat{\mathcal{S}}\), yielding fewer tests and higher power. Moreover, if only one node is truly altered (\(l=1\)) but only pairs are tested, the responsible element of \(M\) cannot be identified.

{
\renewcommand{\thetheorem}{1}
\begin{conjecture}[Behaviour under the alternative in the case of singleton removal]\label{conj:noncentral}
Let \(\mathcal{S}\subset V\) denote the subset of perturbed nodes under \(H_1\); set \(\mathcal{S}^c=V\setminus\mathcal{S}\). 
Assume the regularity conditions of Theorem~\ref{thm:null_distrib_delta_M}. Then, under \(H_1\), as \(n=n_1+n_2\to\infty\),
\[
\Delta_{j}\ \overset{d}{\longrightarrow}\ \chi^2_{\,p+1}(\lambda_{j})
\quad \text{for all } j\in V,
\]
where \(\lambda_j\ge0\) is a node-specific noncentrality parameter. Moreover,
$\lambda_j$ is typically substantially larger for $j\in\mathcal{S}$,
than for $j\in\mathcal{S}^c$.
In particular, nodes in \(\mathcal{S}\) are expected to produce stochastically larger values of \(\Delta_j\), while nodes in \(\mathcal{S}^c\) may still exhibit positive noncentrality due to dependence induced by the covariance structure. In all cases, the degrees of freedom equal \(p+1\), matching the null limit in Corollary~\ref{corollary:null_distrib_delta_j}.
\end{conjecture}
}

Theorem~\ref{thm:null_distrib_delta_M} and Corollary~\ref{corollary:null_distrib_delta_j} provide the asymptotic calibration for valid node or subset-level \(p\)-values under \(H_0\). Simulations in Section~\ref{sec:Simulation study} confirm the accuracy and power of our leave-one-out testing procedure and support Conjecture~\ref{conj:noncentral}.

\section{Simulation study}\label{sec:Simulation study}

\begin{figure}[htbp]
\centering
\includegraphics[width=0.70\textwidth]{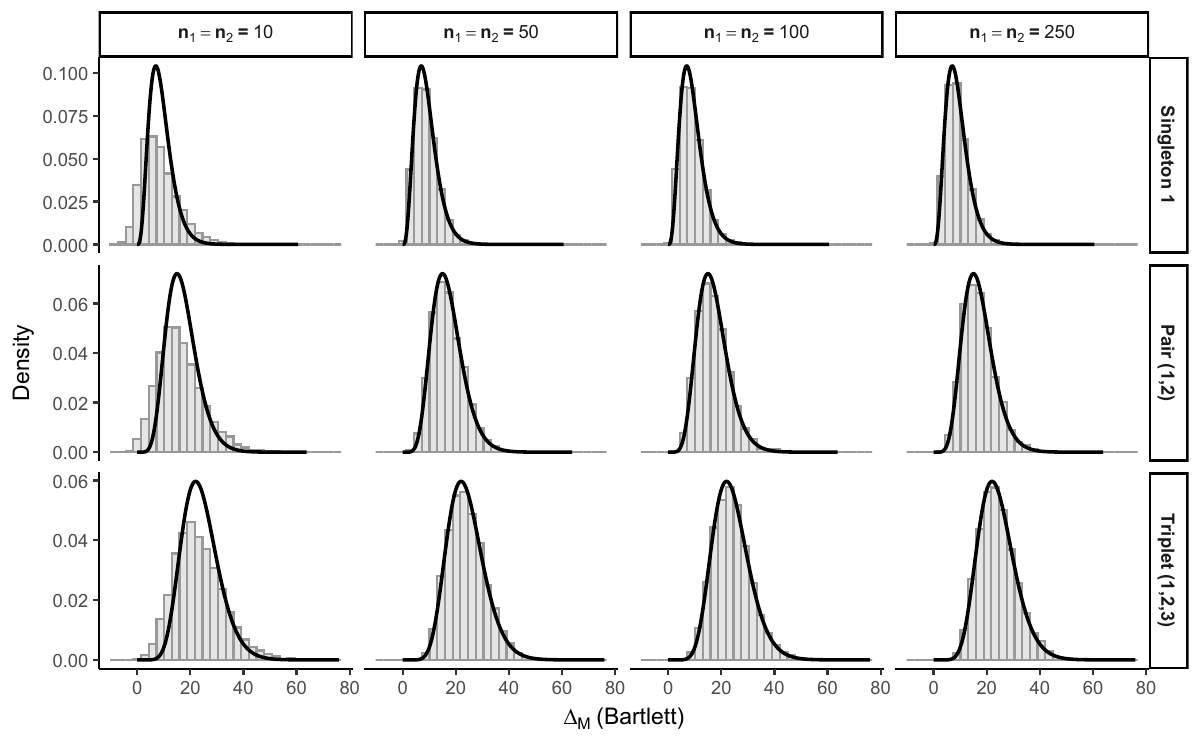}
\caption{Empirical null distribution of Bartlett-adjusted \(\Delta_M\) for \(|M|=l\in\{1,2,3\}\). The black line denotes the theoretical null limiting distribution. Rows vary by \(l\), columns by sample size \(n_1=n_2\in\{10,50,100,250\}\).}
\label{fig:Figure_1}
\end{figure}

\begin{figure}[htbp]
\centering
\includegraphics[width=0.75\textwidth]{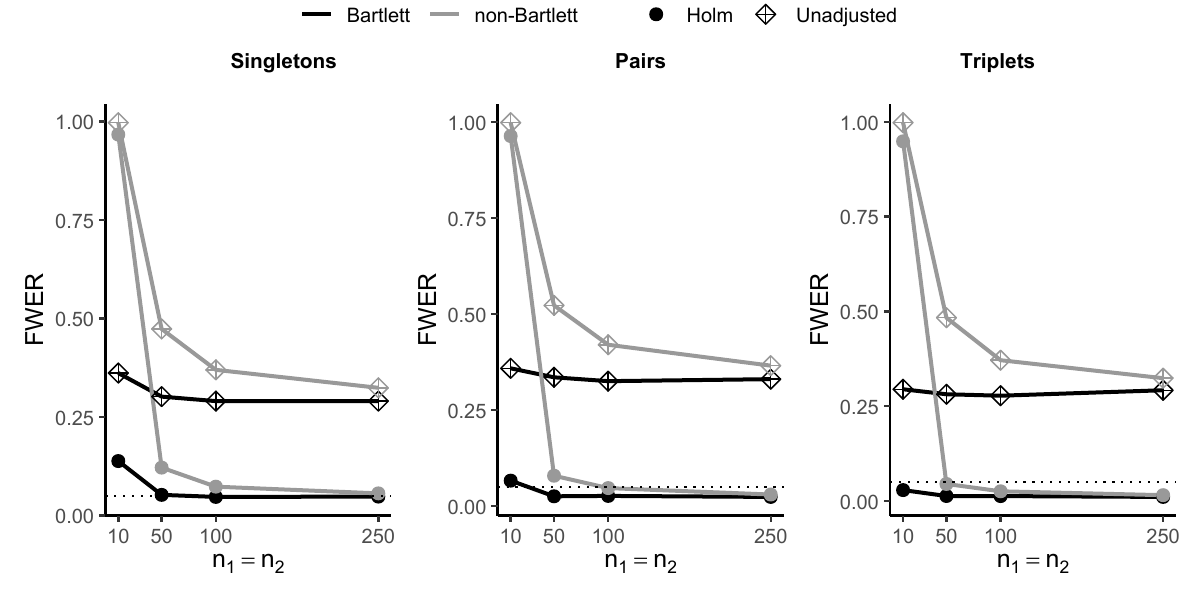}
\caption{FWER under \(H_0\) for singletons ($l=1$), pairs ($l=2$), and triplets ($l=3$), for \(n_1=n_2\). Black and grey lines denote Bartlett and non-Bartlett calibration; point shapes indicate multiplicity adjustment. The horizontal dotted line marks the nominal level \(\alpha=0.05\).
}
\label{fig:Figure_2}
\end{figure}

\begin{figure}[htbp]
\centering
\includegraphics[width=0.65\textwidth]{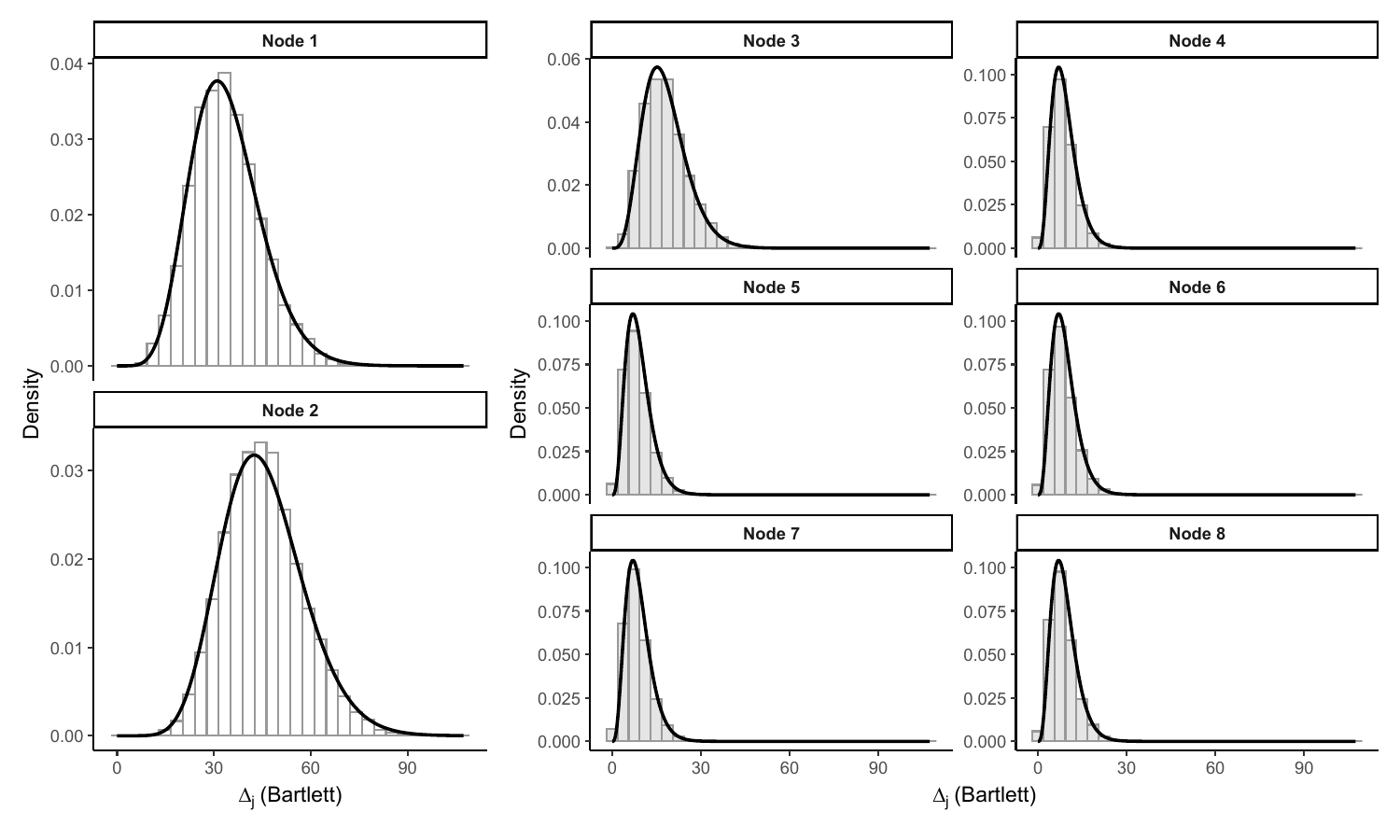}
\caption{Empirical non-null distribution of Bartlett-adjusted \(\Delta_j\) in scenario (S2) with changed nodes \(\mathcal{S}=\{1,2\}\), \(\delta_\mu=1.5\), and \(\xi=0.5\). The black line denotes the theoretical non-null limiting distribution. Results for \(n_1=n_2=100\).  
}
\label{fig:Figure_3}
\end{figure}

We consider 
a fully connected graph with $p = 8$ nodes, interpreted either as the entire graph or as a maximal clique within a larger decomposable structure. For \(c=1,2\), we generate the independent data matrices \(\mathbb{X}_V^{(c)}\) as in Section~\ref{sec:proposal}, with balanced sizes \(n_1=n_2\). 
Under the global null \(H_0\) we set $\mu^{(1)}=\mu^{(2)}=\mu_0=0_p$ and
$\Sigma^{(1)}=\Sigma^{(2)}=\Sigma_0$,
with $\Sigma_0=\Gamma\Gamma^\top+\rho I_p$,
where \(\rho=0.4\), \(I_p\) is the \(p\times p\) identity matrix, and the entries of \(\Gamma\in\mathbb{R}^{p\times p}\) are independently sampled from a standard Gaussian distribution. The matrix $\Sigma_0$ is then standardized to have unit diagonal entries; the corresponding baseline covariance and implied precision matrices are reported in Appendix~A.
Under \(H_1\), fix \(\mathcal S\subset\{1,\ldots,p\}\) with \(|\mathcal S|=k\), and define a mean shift and variance rescaling confined to the nodes in \(\mathcal S\),
\[
\begin{aligned}
\mu^{(2)}_j &=
\begin{cases}
\mu_{0,j} + \delta_\mu,& j\in\mathcal S, \qquad \Sigma^{(2)} = D_\xi\,\Sigma_0\,D_\xi, \\
\mu_{0,j},& j\notin\mathcal S, \qquad D_\xi = \operatorname{diag}(d_1,\ldots,d_p),
\end{cases}
\quad
\begin{aligned}
d_j=
\begin{cases}
\sqrt{\xi},& j\in\mathcal S,\\
1,& j\notin\mathcal S.
\end{cases}
\end{aligned}
\end{aligned}
\]
Thus \(\delta_\mu=0\) and \(\xi=1\) yield the setting under \(H_0\). 
The simulation grid spans \(n_1=n_2\in\{10,50,100,250\}\), \(\delta_\mu\in\{0,0.5,1.0,1.5\}\), \(\xi\in\{0.5,1.0,1.5\}\), and three scenarios for \(\mathcal S\): \(k=1\) with \(\mathcal S=\{1\}\) (S1), \(k=2\) with \(\mathcal S=\{1,2\}\) (S2), and \(k=5\) with \(\mathcal S=\{1,2,3,4,5\}\) (S3), as shown in Fig.~A.1 of  Appendix \ref{app:Appendix_A}. For each configuration we run \(B=5000\) independent simulations. On each dataset we compute the LRT statistic \(W_n^V\) and its Bartlett-adjusted counterpart \(T_n^V\), as well as the increments \(W_n^V-W_n^{V\setminus M}\) and \(T_n^V-T_n^{V\setminus M}\) for all subsets \(M\), with \(|M|=l\in\{1,2,3\}\). Calibration under \(H_0\) uses the \(\chi^2_{h(l,p)}\) limits of Theorem~\ref{thm:null_distrib_delta_M}. 
Fig.~\ref{fig:Figure_1} assess the validity of the null limiting distribution stated in Theorem~\ref{thm:null_distrib_delta_M}, showing adherence of the empirical distributions for arbitrary subsets \(M\) with \(|M|=l\in\{1,2,3\}\) to the expected chi-square limits, with agreement improving as sample size increases. 
Table~\ref{tab:marginal_deltas_grouped} shows marginal type I error control at nominal level \(\alpha=0.05\), for each subset \(M\). 
Bartlett-adjusted increments provide better marginal type I error control than their $W_n$-based counterparts, particularly at small sample sizes; see the leftmost columns for \(n_1=n_2=10\).
Fig.~\ref{fig:Figure_2} shows the empirical FWER under \(H_0\), defined as the proportion of replicates in which at least one null node or subset is falsely rejected. 
In Fig.~A.2 in Appendix \ref{app:Appendix_A} we report power results for $l=1$. Panel~(a) shows the probability of detecting at least one altered node, and panel~(b) the probability of detecting all altered nodes, both versus $\delta_\mu$, at $n_1=n_2\in\{50,100\}$, with Holm correction. Using $l=1$ improves complete recovery when shifts affect few nodes, see scenario (S1), while still ensuring partial recovery when more nodes are perturbed, see scenario (S3).
To study behaviour under \(H_1\) for singletons, we additionally examine the empirical distribution of \(\Delta_j\) and its noncentral chi-square approximation suggested by Conjecture~\ref{conj:noncentral}. For each node we estimate the noncentrality parameter by \(\widehat\lambda_j=\overline{\Delta}_j-(p+1)\), where \(\overline{\Delta}_j\) is the mean of the Bartlett-adjusted singleton increment for node \(j\). Fig.~\ref{fig:Figure_3} compares the empirical distribution of \(\Delta_j\) under \(H_1\) with the distribution suggested by Conjecture~\ref{conj:noncentral} for scenario (S2).

Although the focus of our methodology is on detecting node-level distributional differences arising from mean and variance perturbations, we also explored scenarios in which the difference between the two groups arises from a change in the precision matrix. In particular, we consider localized perturbations affecting a single edge. While this setting is not the primary target of the proposed procedure, it provides insight into how changes in the dependence structure propagate to the node-level statistics. Additional simulations show that such edge perturbations, including edge deletion, can still be detected indirectly through the identification of nodes adjacent to the altered edge; see Fig.~A.17 in Appendix~\ref{app:Appendix_A} and the related discussion.
Further simulation results, together with a real data application to the ALL dataset \citep{Chiaretti_2005,ALL_R_package}, are collected in Appendix~\ref{app:Appendix_A}.
Fig.~A.3 shows the adherence of the global LRT and its Bartlett-adjusted version to their chi-square limits under $H_0$ across sample sizes. Figs.~A.4 and A.5 show well-calibrated singleton increments with Bartlett adjustment and anti-conservative behaviour without it. Figs.~A.6--A.13 extend these diagnostics to pairs and triplets, and Figs.~A.14--A.16 show power curves. 
Figure~A.17 analyzes a scenario in which the perturbation affects the dependence structure rather than the mean or variance, and reports the probability of detecting localized perturbations, including edge deletion, for the pair $(1,2)$, through node-level tests. 
Figs.~A.18--A.24 report simulation results analogous to those of the main study under the misspecified setting in which a sparse graph is treated as fully connected. Despite the discrepancy between the assumed and true graph structures, the proposed procedure exhibits behaviour similar to that observed in the fully connected setting, with effective localization of perturbed nodes, entailing only a mild loss of power due to higher number of tests.
All the simulations were implemented in \textsf{R}; all reproducible code is available at
\url{https://github.com/DavideBenussi/LOOGGM}.

\begin{table}[htbp]
\centering
\footnotesize
\caption{Empirical type I error under $H_0$ for singleton tests ($l=1$) in scenario (S1). Entries are rejection frequencies at $\alpha=0.05$ for the non-Bartlett statistic $\Delta_j$ based on $W_n$ and its Bartlett-adjusted version based on $T_n$, varying $n_1=n_2\in\{10,50,100,250\}$.}
\label{tab:marginal_deltas_grouped}

\begin{tabular}{lllllllll}
\toprule
\multicolumn{1}{c}{ } & \multicolumn{2}{c}{$n_1=n_2=10$} & \multicolumn{2}{c}{$n_1=n_2=50$} & \multicolumn{2}{c}{$n_1=n_2=100$} & \multicolumn{2}{c}{$n_1=n_2=250$} \\
\cmidrule(l{3pt}r{3pt}){2-3}
\cmidrule(l{3pt}r{3pt}){4-5}
\cmidrule(l{3pt}r{3pt}){6-7}
\cmidrule(l{3pt}r{3pt}){8-9}
Node & $\Delta_j\ (W_n)$ & $\Delta_j\ (T_n)$ & $\Delta_j\ (W_n)$ & $\Delta_j\ (T_n)$ & $\Delta_j\ (W_n)$ & $\Delta_j\ (T_n)$ & $\Delta_j\ (W_n)$ & $\Delta_j\ (T_n)$ \\
\midrule
1 & 0.835 & 0.119 & 0.092 & 0.053 & 0.054 & 0.044 & 0.053 & 0.050 \\
2 & 0.843 & 0.138 & 0.119 & 0.070 & 0.068 & 0.048 & 0.060 & 0.056 \\
3 & 0.844 & 0.122 & 0.097 & 0.050 & 0.069 & 0.046 & 0.059 & 0.049 \\
4 & 0.843 & 0.124 & 0.085 & 0.046 & 0.069 & 0.048 & 0.054 & 0.048 \\
5 & 0.815 & 0.119 & 0.107 & 0.055 & 0.062 & 0.044 & 0.055 & 0.050 \\
6 & 0.842 & 0.121 & 0.107 & 0.055 & 0.066 & 0.050 & 0.060 & 0.051 \\
7 & 0.838 & 0.119 & 0.114 & 0.054 & 0.073 & 0.050 & 0.042 & 0.038 \\
8 & 0.851 & 0.116 & 0.102 & 0.059 & 0.076 & 0.051 & 0.059 & 0.052 \\
\bottomrule
\end{tabular}

\end{table}

\section{Extension to decomposable graphical models and scalability to larger graphs}
\label{sec:Extension to decomposable graphical models and scalability to larger graphs}

The methodology developed in Section~\ref{sec:proposal}, and illustrated in the simulation study presented in Section \ref{sec:Simulation study}, is formulated under a moderate-dimensional regime, requiring the number of variables to be small relative to the sample size. Therefore, it is not directly applicable to high-dimensional settings where $p$ is equal or larger than $n$. 
Nevertheless, in large decomposable undirected graphs, inference can still be performed locally within cliques, which constitute low-dimensional components. This allows the proposed procedure to remain applicable even when the overall graph is large and \( p > n \), provided that the dimension of the clique under consideration is smaller than the available sample size in each group. In this section, we show how the leave-one-out strategy extends naturally in this framework.

Let \(G\) be an undirected decomposable graph with maximal cliques \(C_1,\dots,C_K\) ordered to satisfy the running intersection property (RIP; \citealp{lauritzen1996graphical}). Define \(\mathcal{I}=\{2,\ldots,K\}\) and
\[
S_i=C_i\cap C_{i-1}\ (i\in\mathcal{I}),\quad S_1=\varnothing,\qquad
R_i=C_i\setminus C_{i-1}\ (i\in\mathcal{I}),\quad R_1=C_1.
\]
If \(X_V\) factorizes over \(G\) \citep{whittaker2009graphical}, then
\[
f(X_V)
= f(X_{C_1})\prod_{i=2}^K\frac{f(X_{C_i})}{f(X_{S_i})}
= f(X_{R_1})\prod_{i=2}^K f(X_{R_i}\mid X_{S_i}).
\]
Under decomposability, the null hypotheses \eqref{global_H0} is equivalent to
\[
H_0=\bigcap_{i=1}^K H_{0,i},\qquad
H_{0,i}:\ X^{(1)}_{R_i}\mid X^{(1)}_{S_i}\ \overset{d}{=}\ X^{(2)}_{R_i}\mid X^{(2)}_{S_i}.
\]
For any \(A\in\{C_1,\dots,C_K,S_1,\dots,S_K,C_1\mid S_1,\dots,C_K\mid S_K\}\), the LRT is
\[
W_n^{A}
=\sum_{c=1}^2 n_c\log\frac{\det\widehat\Sigma_A}{\det\widehat\Sigma^{(c)}_A},
\]
where \(\widehat\Sigma_A\) is the MLE under \(H_0\), and \(\Sigma^{(c)}_A\) is the MLE for group \(c\). Here, \(W_n^{C_i}\) and \(W_n^{S_i}\) test marginal equality on \(C_i\) and \(S_i\), while \(W_n^{C_i\mid S_i}\) tests equality of the conditional law on \(C_i\mid S_i\). The global LRT decomposes as \citep{DJORDJILOVIC_CHIOGNA_2022}
\[
W_n^{V}=\sum_{i=1}^K\bigl(W_n^{C_i}-W_n^{S_i}\bigr)
= W_n^{C_1}+\sum_{i=2}^K W_n^{C_i\mid S_i},
\]
and under \(H_0\), \(W_n^{C_1}\) and \(W_n^{C_i\mid S_i}\), \(i\in\mathcal{I}\), are asymptotically independent with
\[
W_n^{C_1}\ \overset{d}{\longrightarrow}\ \chi^2_{f_{C_1}},\qquad
W_n^{C_i\mid S_i}\ \overset{d}{\longrightarrow}\ \chi^2_{f_{C_i}-f_{S_i}}.
\]
Finite-sample accuracy is improved by Bartlett corrections \citep{ERIKA_BANZATO_2023}. For \(i\in\mathcal{I}\), set
\[
T_n^{C_1}=\delta_n^{C_1}W_n^{C_1},\quad
\delta_n^{C_1}=\tfrac{f_{C_1}}{-2\mu_n^{C_1}},\qquad
T_n^{C_i\mid S_i}=\delta_n^{C_i\mid S_i}W_n^{C_i\mid S_i},\quad
\delta_n^{C_i\mid S_i}=\tfrac{f_{C_i}-f_{S_i}}{-2\mu_n^{C_i\mid S_i}},
\]
so that \(T_n^{C_1}\) and \(T_n^{C_i\mid S_i}\) admit accurate chi-square approximations with the same degrees of freedom even when \(n\) and \(p\) are unbalanced. 
\cite{salviato2019sourceset} propose an algorithm to identify the clique (or separator) containing the nodes responsible for the change. 
This suggests a natural extension of our procedure in structured settings. 
Indeed, when the underlying graph is decomposable, one can first use a clique-level screening method to identify a subset of cliques likely to contain altered nodes, and then apply the leave-one-out strategy of Section~\ref{sec:proposal} locally within those cliques to achieve node-level localization. 
This yields a two-stage approach, where the first stage performs dimensionality reduction and the second stage refines the inference at the node level.
A formal development of this idea, including sample-splitting to ensure valid post-selection inference and a detailed study of its statistical properties, is beyond the scope of the present paper and is the subject of ongoing research.

\section{Conclusions}\label{sec:Conclusions}

We developed node-level inference for two-sample problems in GGMs by introducing a leave-one-out extension of a Bartlett-adjusted global likelihood ratio test. 
The proposed approach targets localization of general distributional differences between the two groups.
The resulting increments admit standard chi-square null limits, providing calibrated inference for individual nodes and fixed-size subsets even in finite samples.
When the number of altered nodes is unknown, classical multiple testing adjustments yield valid selection of the set of altered nodes. The procedure is computationally light, inherits improved finite sample calibration from the Bartlett adjustment, and performs well in both simulations and real-data analyses.
A limitation of the proposed approach is the reliance on a fully connected working graph. This assumption is reasonable when the underlying structure is unknown and a maximal clique idealization is commonly adopted. However, our method remains robust even when applied to a sparse graph while treating it as fully connected, as demonstrated in Appendix~\ref{app:Appendix_A}. This observation also suggests a natural extension to structured settings, as discussed in Section~\ref{sec:Extension to decomposable graphical models and scalability to larger graphs}.
In particular, when the graph exhibits decomposability, one can first localize the region of departure from the null hypothesis using a valid clique-wise method \citep{salviato2019sourceset, DJORDJILOVIC_CHIOGNA_2022, ERIKA_BANZATO_2023}, and subsequently apply the leave-one-out procedure within the selected cliques to achieve node-level localization. This yields a two-stage strategy, while controlling error rates through appropriate across-stage adjustments \citep{Benjamini_Heller_2008,VanDenBerge_2017}.
A formal treatment of such combined procedures, together with extensions to unbalanced designs, higher dimensions, and robustness beyond normality, is left for future work.

\section*{Acknowledgements}

The authors thank Prof.\ Davide Risso (Department of Statistical Sciences, University of Padova), for his valuable insights during the preparation of this manuscript. We would also like to thank the reviewer and the editor for their valuable comments and suggestions, which have significantly improved the presentation of the material.

\appendix

\section{Supplementary material}\label{app:Appendix_A}

The supplementary material provides the proof of Theorem~\ref{thm:null_distrib_delta_M} and reports the real-data application to the ALL dataset \citep{Chiaretti_2005, ALL_R_package}. It also includes additional simulation results for the fully connected graph considered in the main text, together with an assessment of the procedure's ability to identify nodes incident to the altered edge. Further simulations investigate the case where data are generated from a non-fully connected graph while the procedure is applied under a fully connected working graph assumption, illustrating the robustness of the proposed method.

\section*{Data availability}

The data used for the real-data application are publicly available in the R package \texttt{ALL} \citep{ALL_R_package}:
\url{https://bioconductor.org/packages/ALL}

\bibliographystyle{apalike}
\bibliography{references}

\end{document}